\documentclass[fleqn,usenatbib]{mnras}

\usepackage{newtxtext,newtxmath}
\usepackage{color}

\usepackage[T1]{fontenc}

\usepackage{graphicx}	
\usepackage{amsmath}	

\title[Planck cluster with HAE overdensity at z=2]{A {\it Planck}-selected dusty proto-cluster at $z$$=$2.16 associated with a strong over-density of massive H$\alpha$ emitting galaxies}

\author[Y. Koyama et al.]
{Yusei Koyama,$^{1,2}$\thanks{E-mail: koyama@naoj.org}
Maria del Carmen Polletta,$^{3,4}$
Ichi Tanaka,$^{1}$
Tadayuki Kodama$^{5}$,
\newauthor
Herv\'e Dole$^{6}$,
Genevi\`eve Soucail$^{4}$,
Brenda Frye$^{7}$,
Matthew Lehnert$^{8}$,
Marco Scodeggio$^{3}$
\\
\\
$^{1}$Subaru Telescope, National Astronomical Observatory of Japan, 650 North A'ohoku Place, Hilo, HI 96720, U.S.A.\\
$^{2}$Graduate University for Advanced Studies (SOKENDAI), Osawa 2-21-1, Mitaka, Tokyo 181-8588, Japan\\
$^{3}$INAF -- Istituto di Astrofisica Spaziale e Fisica Cosmica Milano, Via A. Corti 12, 20133 Milano, Italy \\
$^{4}$IRAP, Universit\'e de Toulouse, CNRS, CNES, UPS, Toulouse, France\\
$^{5}$Astronomical Institute, Tohoku University, 63 Aramaki, Aoba-ku, Sendai 980-8578, Japan\\
$^{6}$Universit\'e Paris--Saclay, CNRS, Institut d'astrophysique spatiale, 91405, Orsay, France\\
$^{7}$Department of Astronomy/Steward Observatory, University of Arizona, 933 N Cherry Ave., Tucson, AZ 85721, USA\\
$^{8}$Sorbonne Universit\'e, CNRS UMR 7095, Institut d'Astrophysique de Paris, 98bis bvd Arago, 75014, Paris, France\\
}

\date{Accepted XXX. Received YYY; in original form ZZZ}

\pubyear{2020}

\begin{document}
\label{firstpage}
\pagerange{\pageref{firstpage}--\pageref{lastpage}}
\maketitle

\begin{abstract}
We discovered an over-density of H$\alpha$-emitting galaxies associated with a {\it Planck} compact source in the COSMOS field (PHz~G237.0+42.5) through narrow-band imaging observations with Subaru/MOIRCS. This {\it Planck}-selected dusty proto-cluster at $z=2.16$ has 38 H$\alpha$ emitters including six spectroscopically confirmed galaxies in the observed MOIRCS 4$'$$\times$7$'$ field (corresponding to $\sim$2.0$\times$3.5~Mpc$^2$ in physical scale). We find that massive H$\alpha$ emitters with $\log$$(M_{\star}/M_{\odot})$$>$10.5 are strongly clustered in the core of the proto-cluster (within $\sim$300-kpc from the density peak of the H$\alpha$ emitters). Most of the H$\alpha$ emitters in this proto-cluster lie along the star-forming main sequence using H$\alpha$-based SFR estimates, whilst the cluster total SFR derived by integrating the H$\alpha$-based SFRs is an order of magnitude smaller than those estimated from {\it Planck/Herschel} FIR photometry. Our results suggest that H$\alpha$ is a good observable for detecting moderately star-forming galaxies and tracing the large-scale environment in and around high-$z$ dusty proto-clusters, but there is a possibility that a large fraction of star formation could be obscured by dust and undetected in H$\alpha$ observations. 

\end{abstract}




\section{Introduction}

Within the framework of hierarchical growth of large-scale structures of the Universe, galaxy clusters evolve at intersections of the cosmic web across cosmic time (e.g.\ \citealt{Overzier2016}). Galaxy clusters in the local Universe are dominated by red (quiescent) galaxies with old stellar population, and they are believed to be formed in the early universe at $z\gg 1$ accompanying intense starbursts (e.g.\ \citealt{Bower1992}). Young forming clusters are predicted to be observed as strong overdensities of dusty starbursts (\citealt{Casey2016}; \citealt{Chiang2017}), and therefore it is vital to find such star-bursting proto-clusters at high-$z$ to investigate how the properties of galaxies in today's clusters were put in place. 

A growing number of studies have identified such star-forming (or potentially starbursting) proto-cluster candidates in the early universe with various techniques (e.g.\ \citealt{Hayashi2012}; \citealt{Dannerbauer2014}; \citealt{Wang2016}; \citealt{Oteo2018}; \citealt{Strazzullo2018}; \citealt{Lacaille2019}). 
One approach to detect such short-lived (hence rare) dusty objects at high-$z$ is to use FIR--(sub-)millimeter surveys covering a wide area of the sky. A good example of such dust-selected, highly star-forming systems is the proto-cluster identified around SPT2349--56, the brightest unlensed source from the 2,500-deg$^2$ South Pole Telescope (SPT) survey (\citealt{Miller2018}; \citealt{Hill2020}), where a large number of sub-millimeter galaxies (SMGs) at $z=4.3$ are clustered within a compact region and its total cluster star formation rate (SFR) is estimated to be $\sim$10$^4$~M$_{\odot}$/yr. 

{\it Planck}\footnote{{\it Planck} is a project of the European Space Agency (ESA) with instruments provided by two scientific consortia funded by ESA member states (in particular the lead countries France and Italy), with contributions from NASA (USA), and telescope reflectors provided by a collaboration between ESA and a scientific consortium led and funded by Denmark.} is a very powerful facility for selecting high-$z$ proto-clusters of dusty sources, taking advantage of its all-sky coverage in sub-millimeter (\citealt{Planck2015}; \citealt{Clements2014}; \citealt{Flores-Cacho2016}; \citealt{Greenslade2018}; \citealt{Cheng2019}; \citealt{Kubo2019}). Using the {\it Planck} high-$z$ source candidates (PHz) catalogue (\citealt{Planck2016}), in combination with {\it Herschel} photometry (HerMES; \citealt{Oliver2012}), we investigate the region around a {\it Planck} source (PHz~G237.0+42.5) lying in the COSMOS field. Within the 4.5-arcmin {\it Planck} beam (in diameter), there are several {\it Herschel} FIR sources and X-ray sources. Medium-resolution spectroscopy from follow-up NIR LUCI/LBT observations (covering the {\it Planck} beam area with its 4$'$$\times$4$'$ FoV) and optical VIMOS/VLT spectra from the zCOSMOS survey (\citealt{Lilly2007}) reveal 8 sources at $z$$=$2.16 within the {\it Planck} beam area (\citealt{Polletta2020}).

In this Letter, we present H$\alpha$ imaging observations of the PHz~G237 field with Subaru/MOIRCS using a narrow-band (NB) filter (NB2071), which is perfectly matched to the H$\alpha$ lines from the redshift of this structure ($z=2.16$). Narrow-band H$\alpha$ imaging observations are shown to be successful in detecting high-$z$ star-forming galaxies within a narrow redshift slice in various environments (e.g.\ \citealt{Geach2008}; \citealt{Sobral2013}; \citealt{Tadaki2013}; \citealt{Hayashi2016}; \citealt{Darvish2020}). This study presents the first attempt to perform H$\alpha$ imaging observations towards a {\it Planck}-selected proto-cluster---with the aim to reveal the structures traced by typical SF galaxies, and to study the nature of member galaxies residing in the dust-selected proto-cluster at the peak epoch of galaxy formation. Throughout this Letter, we adopt the standard cosmology with $\Omega_M =0.3$, $\Omega_{\Lambda} =0.7$, and $H_0 =70$\,km\,s$^{-1}$\,Mpc$^{-1}$. All magnitudes are given in the AB system, and we assume the \citet{Chabrier2003} initial mass function (IMF) throughout this work.

\section{Observations and H$\alpha$ emitter selection}

We carried out MOIRCS/Subaru observations of the PHz~G237 field using NB2071 filter ($\lambda_c$$=$2.068~$\mu$m, $\Delta$$\lambda$$=$0.027$\mu$m; corresponding to the H$\alpha$ line at $z$$=$2.13--2.17) on December 21, 2018. The observations were executed in service mode (S18B-206S, PI: Y.~Koyama) under very good seeing conditions (FWHM=0.4$''$), and the total exposure time was 180-min. The data are reduced in a standard manner using the {\sc mcsred2} software (\citealt{Tanaka2011}).

We use ULTRA-VISTA $K_s$-band images (\citealt{McCracken2012}) to measure the continuum levels for the NB data. To reliably select emission-line objects, it is important to perform photometry in a consistent way for broad- and narrow-band data. After smoothing our NB image to match the PSF size of the $K_s$-band image (0.8$''$), we use the dual-image mode of SExtractor (ver.2.19.5; \citealt{Bertin1996}) to create a NB-selected source catalogue.  
We use 1.6$''$ aperture photometry (2$\times$PSF size in diameter) to determine the source detection and $K_s$$-$NB colours. By distributing apertures (with the same size) at random positions on the NB and $K_s$ images, we estimate 5$\sigma$ limiting magnitudes for NB2071 and $K_s$ image as 23.46 mag and 25.05~mag, respectively. We apply an aperture correction of 0.34-mag to derive their total magnitudes, based on the median difference between the aperture magnitudes ({\sc mag\_aper}) and petrosian magnitudes ({\sc mag\_auto}).\footnote{Three H$\alpha$ emitters selected in this study do not satisfy $K_s$$-$NB$>$0.25 if we use {\sc mag\_auto} to define a NB excess. The cluster total H$\alpha$ SFR would be reduced by $\sim$12\% if we remove these sources from the analyses.}
We detect 829 NB sources ($>$5$\sigma$), in which 800 (97\%) sources have counterparts in the COSMOS2015 catalogue (\citealt{Laigle2016}). We use those 800 galaxies as the parent sample of this study, out of which 142 sources (18\%) have redshifts from the COSMOS spectroscopic surveys.

\begin{figure}
  \vspace{-2mm}
	\includegraphics[width=8.3cm]{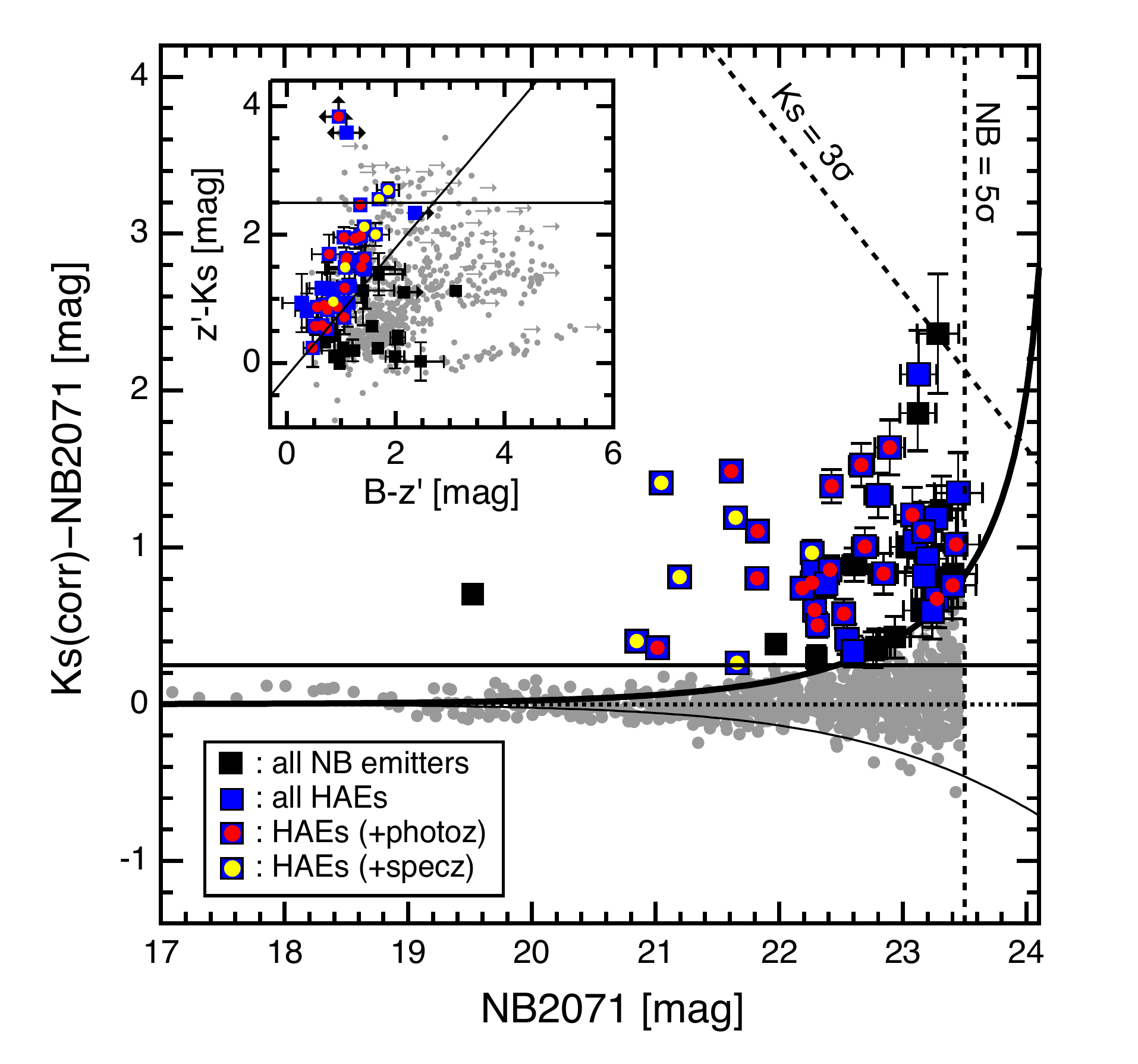}
        \vspace{-4mm}
    \caption{$K_s$$-$NB2071 versus NB2071 magnitudes for all objects in our MOIRCS field (grey dots). The solid-line curves indicate $\pm$2.5$\Sigma$ excess for $K_s$$-$NB colour. 53 galaxies satisfying $K_s$$-$NB$>$0.25 (equivalently EW$_{\rm rest}$$>$30\AA) and $K_s$$-$NB$>$2.5$\Sigma$ are defined as NB emitters (black squares). The blue squares are HAEs selected by spec-$z$ (yellow circles), photo-$z$ (red circles), or {\it BzK} colours (solid blue squares). In the inset, we show the {\it BzK} diagram for all objects, where undetected sources are replaced with their 2$\sigma$ limits and shown as their upper/lower limits. The solid lines indicate the original {\it BzK} selection determined by \citet{Daddi2004}.}
    \label{fig:HAE_select}
\end{figure}

In Fig.~\ref{fig:HAE_select}, we plot $K_s$$-$NB colours against their NB magnitudes. We here applied $-$0.08-mag offset to the $K_s$-band magnitudes to account for the spectral slope of the standard star (GD153) used for NB2071 photometry. We also applied a small offset ($+$0.03-mag) in the NB photometry to set $K_s$$-$NB$=$0 (median) at the bright end, to properly evaluate the NB excess with respect to the continuum flux density at the same wavelength. We define 53 sources which satisfy $K_s$$-$NB$>$0.25 and $K_s$$-$NB$>$2.5$\Sigma$ as NB emitters (black squares in Fig.~\ref{fig:HAE_select}), where $\Sigma$ represents the colour excess in $K_s$$-$NB (e.g.\ \citealt{Bunker1995}). We note that all the NB emitters identified here are also detected at $\gtrsim$3$\sigma$ levels in the ULTRA-VISTA $K_s$-band data. 

To select H$\alpha$ emitters (HAEs) at $z$$=$2.16 and remove interlopers, we use spec-$z$, photo-$z$ (from COSMOS2015 catalogue), and {\it BzK} colours. We first select 6 emitters with 2.150$<$$z_{\rm spec}$$<$2.164 (\citealt{Polletta2020}) as spec-$z$ HAEs. For those without spec-$z$ information, we select HAEs using the photo-$z$ (1.8$<$$z_{\rm photo}$$<$2.4); we here use {\sc zminchi2} from the COSMOS2015 catalogue, defined as the minimum of the $\chi^2$ distribution from the template fitting (see \citealt{Laigle2016} for details). In addition, we select NB emitters satisfying the {\it BzK} criteria (Fig.~\ref{fig:HAE_select}), regardless of their photo-$z$. Because the photo-$z$ uncertainty is often large for SF galaxies with blue/featureless SEDs at $z$$\sim$2, the {\it BzK} selection {\it combined with} the NB excess is shown to be an efficient way for selecting HAEs at the targeted redshifts (e.g.\ \citealt{Koyama2013a}; \citealt{Shimakawa2018}). Amongst the 53 NB emitters, we select 38 HAEs in total; i.e.\ 6 galaxies as spec-$z$ HAEs, 25 galaxies as photo-$z$ HAEs (5 of which are spec-$z$ HAEs), and 12 additional galaxies as {\it BzK} HAEs. As can be seen in the {\it BzK} diagram, most of the spec-$z$ and photo-$z$ HAEs also satisfy {\it BzK} selection. 
We note that three NB emitters are not detected at $B$- or $z'$-band, one of which is undetected at both $B$- and $z'$-band. We keep these sources as their lower/upper limits still satisfy the {\it BzK} criteria. We note that the source undetected at both $B$- and $z'$-band is indicated with "{\sf B}" mark in Fig.~\ref{fig:map}, and this source contributes 4.5\% of the cluster total SFR. 

\section{Results and Discussion}

\subsection{Massive H$\alpha$ emitters in the proto-cluster core}
\label{sec:spatial_distribution}

\begin{figure}
	\includegraphics[width=8.7cm]{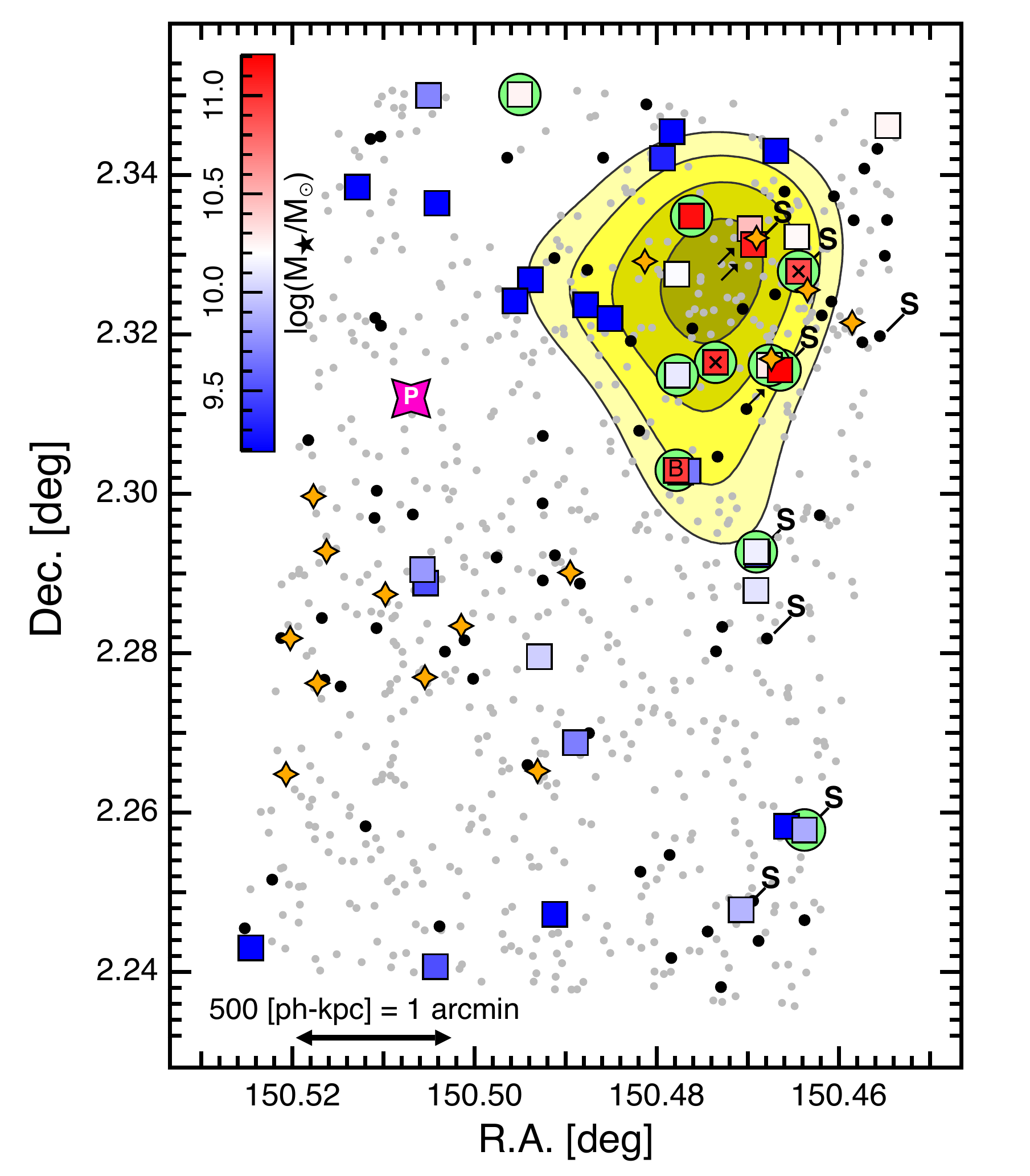}
        \vspace{-5mm}
    \caption{The 2-D sky distribution of HAEs (coloured squares), with the colour code indicating their $M_{\star}$. The grey circles show all NB sources, out of which black circles indicate galaxies with 1.8$<$$z_{\rm photo}$$<$2.4. The green circles are 24$\mu$m-detected HAEs. Two sources with ``{\sf x}'' mark are X-ray detected HAEs, and eight sources connected to ``{\sf S}'' marks are spectroscopic members. Three arrows indicate HAEs which do not satisfy $K_s$$-$NB$>$0.25 in the case we use {\sc mag\_auto}, and one source with ``{\sf B}'' mark shows the HAE which is undetected in both $B$- and $z'$-bands. To guide the eye, the yellow-shaded contours indicate 1.5, 2.0, 2.5, 3.0-$\sigma$ above the median local number density measured with HAEs. The orange stars show the red {\it Herschel} sources (see Section~3.3). The coordinate of the {\it Planck} source (R.A., Dec.)$=$(150.507, $+$2.31204) is shown with the pink star with ``{\sf P}'' mark.} 
    \label{fig:map}
\end{figure}

We show in Fig.~\ref{fig:map} the 2-D distribution of the HAEs (squares), photo-$z$ selected potential cluster members (1.8$\le$$z_{\rm photo}$$\le$2.4; black circles), and all the NB-detected objects (grey circles) in the PHz~G237 field. Objects with ``{\sf S}'' marks are spectroscopic members, six of which are HAEs. We compute the local number density of HAEs at a given point by applying gaussian smoothing ($\sigma$$=$300~kpc) for each HAE and by combining the tails of those gaussian wings (see contours in Fig.~\ref{fig:map}). The smoothing radius is chosen following the typical size of HAE overdensities associated with rich proto-clusters at similar redshifts (e.g.\ \citealt{Koyama2013a}). We determine (R.A., Dec.)$=$(10:01:53.67, $+$02:19:38.9) as the HAE density peak. 

In Fig.~\ref{fig:map}, HAE symbols are colour coded based on their $M_{\star}$; i.e.\ redder colours indicate higher $M_{\star}$. Stellar masses of HAEs are derived by fitting the SEDs ({\it uBVrizJHK} and IRAC ch1 and ch2 photometry from COSMOS), using the {\sc fast} code (\citealt{Kriek2009}). We here assume the fixed redshift ($z$$=$2.16), the \citet{Bruzual2003} stellar population synthesis model, the \citet{Calzetti2000} dust attenuation law, and the \citet{Chabrier2003} IMF. We also assume exponentially declining SFRs (SFR$\propto\exp(-t/\tau)$) with parameter ranges of $\tau$ and age of 10$^7$--10$^{10}$~yr, metallicity$=$[0.004, 0.008, 0.02], and $A_V$$=$0.0--3.0~mag. 

Fig.~\ref{fig:map} demonstrates that massive HAEs (with $M_{\star}$$\gtrsim$10$^{10.5}M_{\odot}$) are strongly clustered around the density peak. A similar trend was reported for other candidate proto-clusters at $z$$\gtrsim$2 (e.g.\ \citealt{Hatch2011}; \citealt{Matsuda2011}; \citealt{Koyama2013a}), suggesting an accelerated galaxy growth in dense environments in the early universe. We note that many of those massive HAEs residing in the proto-cluster environment are individually detected at Spitzer/MIPS 24$\mu$m image (as shown with green circles in Fig.~\ref{fig:map}). Their stellar masses are already comparable to the present-day massive cluster galaxies, but they are still actively forming stars (see also Section~\ref{sec:sfms}). It should also be noted that two of the massive HAEs near the density peak are X-ray sources (see ``{\sf x}'' marks in Fig.~\ref{fig:map}), suggesting H$\alpha$ emission of these two galaxies may be contributed by AGNs. The stellar masses of these X-ray HAEs may be overestimated, but their stellar masses would still be higher than the median $M_{\star}$ of our HAEs sample, even if we consider a significant fraction of their $M_{\star}$ estimates are contributed by AGNs (see Section~\ref{sec:sfms}).

\subsection{Proto-cluster HAEs on the SFR--$M_{\star}$ relation}
\label{sec:sfms}

We derive the H$\alpha$+[N{\sc ii}] line fluxes ($F_{\rm{H\alpha+[NII]}}$), continuum flux density ($f_c$), and the rest-frame equivalent widths (EW$_{\rm rest}$) of HAEs from the $K_s$-band and NB2071 photometry in the same way as \citet{Koyama2013a}. We estimate the contribution of [N{\sc ii}] lines ([N{\sc ii}]/H$\alpha$ ratio) and H$\alpha$ dust attenuation ($A_{\rm H\alpha}$) using the empirical calibrations established for local SF galaxies; we use [N{\sc ii}]/H$\alpha$--EW$_{\rm rest}$(H$\alpha$) relation presented by \citet{Sobral2012}, and the $A_{\rm H\alpha}$--$M_{\star}$--EW$_{\rm rest}$(H$\alpha$) relation from \citet{Koyama2015}\footnote{We confirmed that our conclusions are unchanged even if we use the $A_{V}$ derived from SED fitting to predict $A_{\rm H\alpha}$ (assuming the \citet{Calzetti2000} curve and $E$(B$-$V)$_{\rm star}$/$E$(B$-$V)$_{\rm gas}$$=$0.44). We find that the average dust-corrected SFRs could be higher by a factor of $\sim$1.8$\times$ in this case, but we caution that there is a large uncertainty in the $E$(B$-$V)$_{\rm star}$/$E$(B$-$V)$_{\rm gas}$ ratio (which can be 0.44--1.0; see e.g. \citealt{Koyama2019}).}. We then convert the H$\alpha$ luminosity to SFR$_{\rm H\alpha}$ using the \cite{Kennicutt1998} relation by taking into account the IMF difference. 

\begin{figure}
  \vspace{-2mm}
  \hspace{-3mm}
	\includegraphics[width=9.2cm]{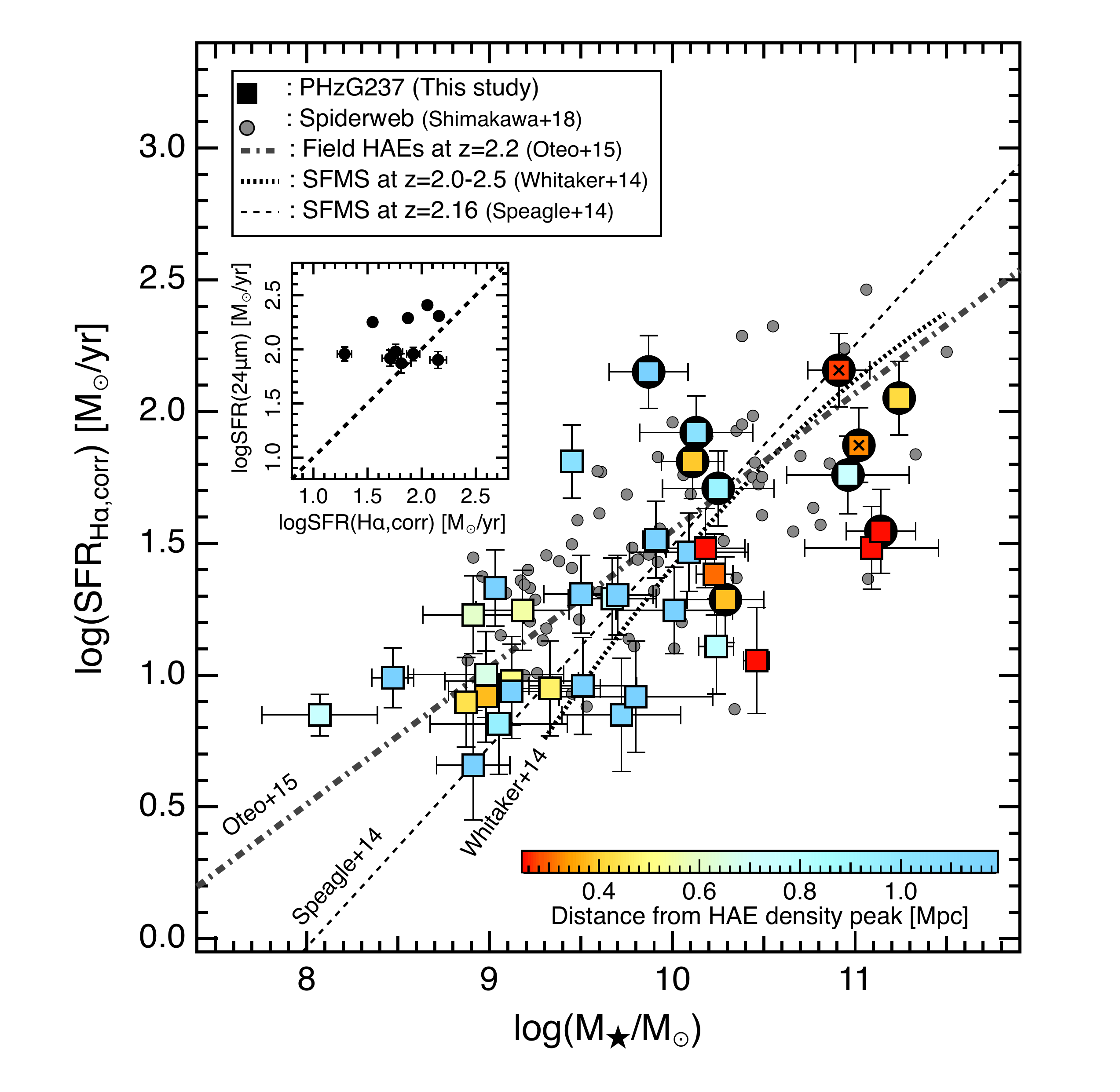}
        \vspace{-8mm}
    \caption{SFR--$M_{\star}$ diagram for HAEs in the PHz~G237 proto-cluster field (squares). The colour coding indicates the distance from the HAE density peak. The large black circles show 24$\mu$m-detected HAEs, and the ``{\sf x}'' marks indicate X-ray sources. We also show HAEs in the Spiderweb proto-cluster at the same redshift from \citet{Shimakawa2018}, and the dot-dashed line indicates the SFR--$M_{\star}$ relation for HAEs in field environment at $z\sim 2$ (\citealt{Oteo2015}). For comparison, we also show the SFMS at $z\sim 2$ defined by \citet{Speagle2014} and \citet{Whitaker2014}. In the inset, we compare the SFRs derived from H$\alpha$ and 24$\mu$m fluxes for 24$\mu$m-detected HAEs.}
    \label{fig:SFMS}
\end{figure}

In Fig.~\ref{fig:SFMS}, we show the SFR--$M_{\star}$ diagram for the HAEs in the PHz~G237 field (coloured symbols). In this plot, the colour coding indicates the distance from the HAE density peak; the redder colour symbols show HAEs located closer to the highest-density region. It can be seen that the $M_{\star}$ distribution of HAEs near the density peak (red/orange symbols) tend to be skewed to the massive end, as already discussed in Section~\ref{sec:spatial_distribution}. We also plot in Fig.~\ref{fig:SFMS} the HAEs in the Spiderweb proto-cluster at the same redshift ($z$$=$2.16) selected using the same NB filter and the same instrument (\citealt{Shimakawa2018}; see also \citealt{Koyama2013a}). For HAEs with $\log(M_{\star}/M_{\odot})$$>$9.5, we find that the fraction of massive HAEs (with $\log(M_{\star}/M_{\odot})$$>$10.5) is consistent between PHz~G237 (26$\pm$12\%) and Spiderweb (30$\pm$9\%).
The dot-dashed line in Fig.~\ref{fig:SFMS} is the SF main sequence (SFMS) derived for the field HAEs at $z$$\sim$2 (\citealt{Oteo2015}). We note that all these studies are based on H$\alpha$ emitters, and use similar approaches to derive $M_{\star}$ and SFRs. Our results suggest that the SFMS for HAEs in the PHz~G237 proto-cluster field are broadly consistent with those in the Spiderweb proto-cluster and in the general field at similar redshifts, suggesting similar mass growth rates in all environments at a given stellar mass, consistent with our previous studies (e.g.\ \citealt{Koyama2013b}). 

The two X-ray detected HAEs are shown with "{\sf X}" marks in Fig.~\ref{fig:SFMS}. If we remove these two AGN candidates, massive HAEs within $\sim$500-kpc from the density peak tend to be located {\it below} the SFMS. This result may suggest a decline of specific SFRs in dense environments, but we must wait for future spectroscopic observations because accurate dust attenuation correction is critical to reliably measure the SFRs of such massive SF galaxies at high-$z$. In fact, there are two 24$\mu$m-detected HAEs located significantly below the SFMS in Fig.~\ref{fig:SFMS}, but their SFRs derived from 24$\mu$m photometry (using the SED templates presented by \citealt{Wuyts2008}) turn out to be $\sim$0.6--0.7-dex higher than what we expected from the dust-corrected H$\alpha$ line (see the inset in Fig.~\ref{fig:SFMS}). Unfortunately, the IR-based SFRs are not available for other MIPS-undetected HAEs (and SFRs from 24$\mu$m alone also suffer from large uncertainties). In any case, we need spectroscopic follow-up observations to accurately measure the dust attenuation effects, to discuss the exact locations of proto-cluster HAEs with respect to the SFMS.  

\subsection{Cluster total SFR from H$\alpha$ and IR}
\label{sec:cluster_sfr}

\begin{figure}
	\includegraphics[width=8.6cm]{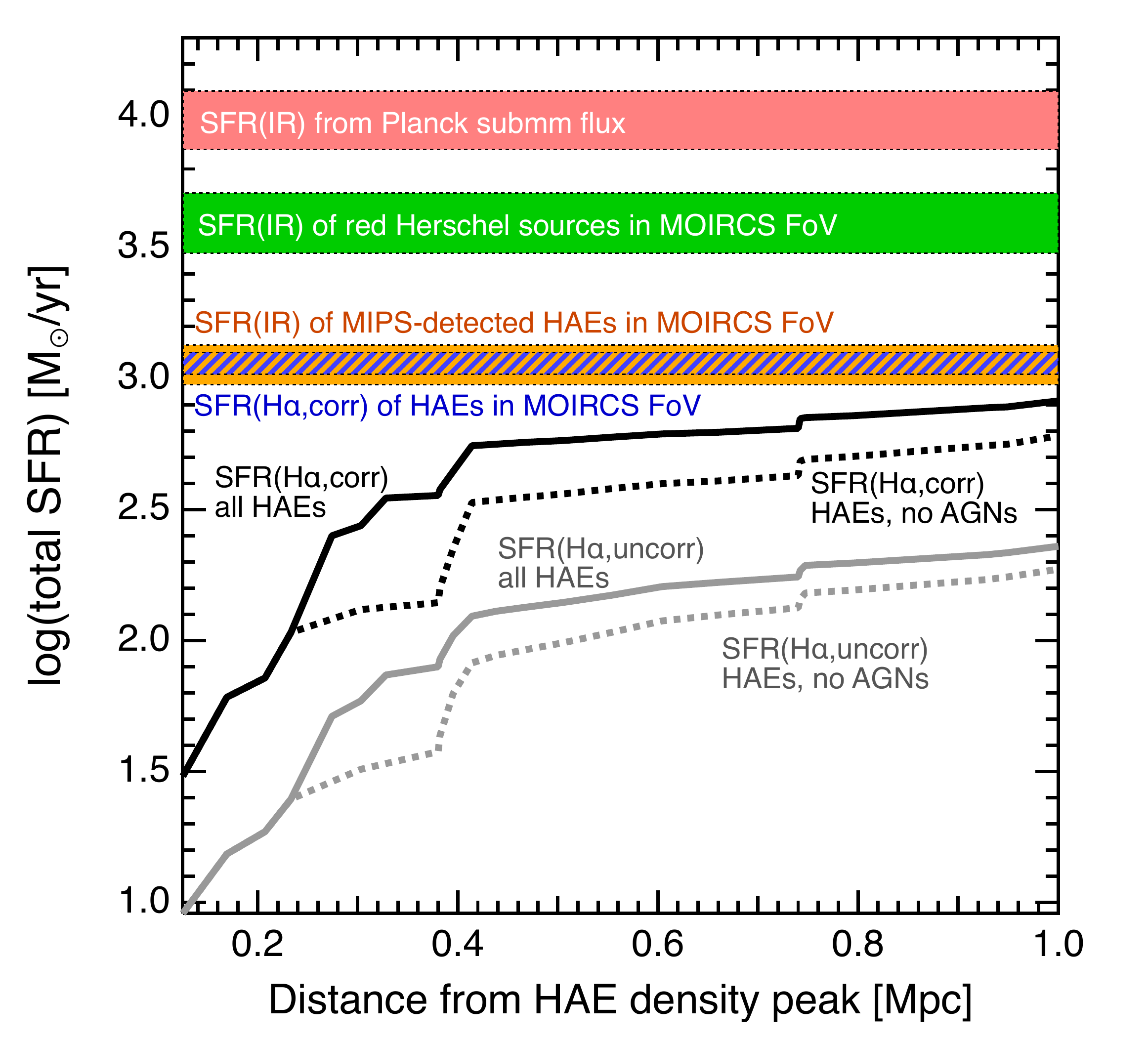}
        \vspace{-5mm}
    \caption{Cluster-integrated (cumulative) SFRs for HAEs in the PHz~G237 proto-cluster field plotted against the distance from the HAE density peak. The black/grey lines show the results for the H$\alpha$-based SFRs with/without dust attenuation correction. In both cases, solid lines are the results for all HAEs, while the dashed lines are the results for non-AGN HAEs by removing two X-ray sources. We also show the cluster total SFRs measured with various methods. The blue hatched area shows the total dust-corrected SFR(H$\alpha$)s for all 38 HAEs within MOIRCS FoV, while the orange shaded region shows the sum of the IR-derived SFRs for 24$\mu$m-detected HAEs in the same FoV (the widths of the stripes indicate the results with/without the AGN effects). We also show the FIR-derived total SFR measured from {\it Planck} sub-millimeter flux (red shaded region) and from FIR sources with red {\it Herschel} colours located within the MOIRCS FoV (green shaded region). 
}
    \label{fig:total_sfr}
\end{figure}

In Fig.~\ref{fig:total_sfr}, we show the cumulative SFR distribution from the HAE density peak (out to 1-Mpc). The black and grey lines represent the results when we use dust-corrected and dust-uncorrected H$\alpha$ SFRs, respectively. For both black and grey lines, the solid lines show the results for all HAEs, while the dashed lines show the results when we remove two X-ray AGNs. We also show in Fig.~\ref{fig:total_sfr} the cluster-integrated SFRs measured with various approaches. The blue hatched region shows the integrated dust-corrected SFR$_{\rm H\alpha}$ for all HAEs within the MOIRCS FoV (the width of the hatched region indicates the results with/without AGNs). The orange shaded region shows the integrated SFR(MIR)s of the 24$\mu$m-detected HAEs. It seems that the total dust-corrected SFR$_{\rm H\alpha}$ are consistent with 24$\mu$m-based SFRs, but we caution that the total MIPS SFR is computed only for 10 HAEs which are individually detected at 24$\mu$m, and thus can be regarded as the lower limit for the cluster total SFR. 

The green shaded region in Fig.~\ref{fig:total_sfr} shows the cluster total SFR$_{\rm FIR}$ as a sum of 15 {\it Herschel} sources with red FIR colours ($S_{350}/S_{250}$$>$0.7 and $S_{500}/S_{350}$$>$0.6; \citealt{Planck2015}) in the MOIRCS FoV (see orange stars in Fig.~\ref{fig:map}). Here we use {\sc cmcirsed} package (\citealt{Casey2012}) to fit the {\it Herschel}/SPIRE photometry with a modified grey-body function at $T_{\rm dust}$$=$30~K and $T_{\rm dust}$$=$35~K (with fixed $\beta$$=$1.8) to estimate the total $L_{\rm IR}$(8--1000$\mu$m). The FIR colour criteria are supposed to select galaxies at $z$$\sim$2--4 (and may suffer from fore-/background interlopers), but it is worth noting that the total SFR could be $\sim$5$\times$ higher than what we expect from H$\alpha$. We also calculate the SFR(FIR) from the {\it Planck} 857GHz and 545GHz fluxes (assuming $z=2.16$ and $T_{\rm dust}$$=$30~K; the red shaded region in Fig.~\ref{fig:total_sfr}). We must regard this as an upper limit of the cluster total SFR as the {\it Planck} fluxes are contributed by all sources within the {\it Planck} beam (see consistent results in \citealt{Planck2015}). We also note that our H$\alpha$ data does not cover the full {\it Planck} beam area (see Fig.~\ref{fig:map}). In addition, some sources may lie outside the redshift range covered by the NB filter ($\Delta$$v$$\gtrsim$2,000~km/s). Therefore, we cannot rule out the possibility that there exist other structures contributing to the {\it Planck} fluxes outside the current H$\alpha$ survey.

Our results suggest that the cluster total SFR measured with H$\alpha$ could be an order of magnitude smaller than IR-based SFRs. This large difference may in part come from uncertainties in calibration between H$\alpha$- and IR-based SFRs at high-$z$, or we may underestimate the H$\alpha$ dust attenuation for individual HAEs. It is also possible that a large number of cluster members which are not identified as HAEs contribute to the cluster total IR luminosities (like ``optically dark'' galaxies; e.g.\ \textcolor{red}{\citealt{Simpson2014}}; \citealt{Franco2018}). Deep and wide-field sub-millimeter mapping with ALMA (e.g.\ \citealt{Umehata2017}; \citealt{Kneissl2019}) would be a powerful approach to resolve the dust-enshrouded SF activity within young dusty proto-clusters.  

\vspace{-5mm}
\section{Summary}
\label{sec:summary}

With our Subaru/MOIRCS NB H$\alpha$ imaging observations towards a {\it Planck} compact source lying in the COSMOS field (PHz~G237.0+42.5), we reported the discovery of a dusty proto-cluster at $z=2.16$ associated with an over-density of massive HAEs. We identified 38 HAEs at $z=2.16$ within the observed 4$'$$\times$7$'$ FoV, out of which 6 galaxies are spectroscopically confirmed. We find that massive HAEs (with $M_{\star}$$>$10$^{10.5}M_{\odot}$) are strongly clustered in the proto-cluster core region (within $\sim$300-kpc from the HAE density peak). Most of the HAEs in the PHz~G237 proto-cluster region are typical SF galaxies on the SFMS. While we find the most massive HAEs may have suppressed SFRs (from H$\alpha$-based SFR estimates), their 24$\mu$m-derived SFRs are consistent with lying along the SFMS. By integrating the H$\alpha$-based SFRs of all H$\alpha$ emitters, we estimate that the cluster total H$\alpha$ SFR could be an order of magnitude smaller than those predicted from {\it Planck/Herschel} FIR photometry. Our results suggest that H$\alpha$ is a good indicator for detecting moderately SF galaxies and tracing the large-scale environment in and around high-$z$ dusty clusters, but there remains a possibility that a significant amount of star formation is obscured by dust and unseen by H$\alpha$ observations. 

\vspace{-3mm}
\section*{Acknowledgements}

We thank the reviewer for their careful read and constructive comments, which improved the paper. The narrow-band imaging data used in this paper are collected at the Subaru Telescope, which is operated by the National Astronomical Observatory of Japan (NAOJ). This work was financially supported in part by a Grant-in-Aid for the Scientific Research (No.18K13588) by the Japanese Ministry of Education, Culture, Sports and Science. MP acknowledges the financial support from Labex OCEVU.  This research has made use of data from HerMES project (http://hermes.sussex.ac.uk/).

\vspace{-3mm}
\section*{Data Availability}

The data underlying this article will be shared on reasonable request to the corresponding author.







%
%
%
%

\bsp	
\label{lastpage}
\end{document}